\newcommand{\be}{\begin{equation}}
\newcommand{\ee}{\end{equation}}
\newcommand{\bea}{\begin{eqnarray}}
\newcommand{\eea}{\end{eqnarray}}
\newcommand{\de}{\partial}
\newcommand{\nn}{\nonumber}
\newcommand{\bra}[1]{\big{<}#1\big{|}}
\newcommand{\ket}[1]{\big{|}#1\big{>}}

\newcommand{\da}{\dagger}
\newcommand{\ep}{\epsilon}
\newcommand{\epl}{\lim _{\epsilon \rightarrow 0}} 
\newcommand{\pa}[1]{\left( #1\right)}
\newcommand{\psb}{\Psi _{base}}
\newcommand{\he}{\hat E}

\newcommand{\hr}{\hat \rho}
\newcommand{\eh}{e \hbar}
\newcommand{\gt}{\tilde \gamma}
\newcommand{\lone}{\hat L_{\alpha \alpha}}
\newcommand{\lgn}{\gamma _{\alpha \beta}^0 \hat L_{\alpha \beta}}
\newcommand{\lgo}{\gamma _{\alpha \beta}^1 \hat L_{\alpha \beta}}
\newcommand{\lgg}{\gamma _{\alpha \beta} \hat L_{\alpha \beta}}
\newcommand{\ppl}{\Psi ^+_{(n,x_1,y_1,\ldots ,x_a,y_a)}}
\newcommand{\pmi}{\Psi ^-_{(n,x_1,y_1,\ldots ,x_a,y_a)}}
\newcommand{\pplo}{\Psi ^+_{(n_0,x_1,y_1,\ldots ,x_a,y_a)}}
\newcommand{\pmio}{\Psi ^-_{(n_0,x_1,y_1,\ldots ,x_a,y_a)}}
\newcommand{\leave}{\! \! \! \! \! / \, \,}
\newcommand{\refpa}[1]{(\ref{#1})}
\newcommand{\intl}[1]{\int d\! #1 \,} 
\newcommand{\lp}{\hat L_+}
\newcommand{\lm}{\hat L_-}
\newcommand{\kp}{\hat K_+}
\newcommand{\km}{\hat K_-}

\documentstyle [12pt]{article}
\begin{document}
\begin{titlepage}
\begin{flushright}
Revised version
\end{flushright}
\begin{flushleft}
G\"oteborg\\
ITP 93-8\\
hep-th/9304101\\
May 1993\\
\end{flushleft}
\vspace{1cm}
\begin{center}
{\Large QED$_{1+1}$ by Dirac Quantization}\\
\vspace{5mm}
{\large Joakim Hallin}\footnote{Email address: tfejh@fy.chalmers.se}\\
\vspace{1cm}
{\sl Institute of Theoretical Physics\\
Chalmers University of Technology\\
and University of G\"oteborg\\
S-412 96 G\"oteborg, Sweden}\\
\vspace{1.5cm}
{\bf Abstract}\\
\end{center}
Dirac quantization of electrodynamics on a two-dimensional cylindrical
space-time, is considered in an explicit loop and functional representation.
Expressions for a sufficient number of physical states are found, so that an
inner product can be derived on the space of physical states. The
relevant states are found to be a superposition of two disjoint sectors. The
ground state is investigated and the investigations are met with
partial success.
\end{titlepage}

\section{Introduction and motivation}
One of the main approaches towards a consistent quantization of gravity is
Dirac quantization of general relativity formulated in
Ashtekar's variables \cite{aa:lect}. However, it appears to us that Dirac
quantization of genuine, interacting  field theories is a poorly
investigated subject. This paper tries to fill that gap by considering a
non-trivial model, electrodynamics on a $1+1$-dimensional
space-time, which can be thought of as a model for gravity coupled to fermions
at the kinematical level. It can of course also be thought of
as a model for electrodynamics or Yang-Mills coupled to fermions in
$3+1$-dimensions. The reason for choosing this particular model is that
in the case of massless fermions, it was shown by Schwinger to be exactly
solvable and hence the massless case is generally known as the
Schwinger model. This model has been investigated in numerous papers over the
years. Schwinger, however, considered space to be a real line. The
 Schwinger model on a circle has been considered in \cite{nm:schw} and
\cite{hh:circle}. One of the crucial ingredients in the quantization of general
relativity is
the imposition of the so called ``reality conditions'', which ensures that the
classical limit of the quantum theory has the correct reality
properties. In the Ashtekar quantization program, the reality conditions are
hoped to pick out the correct inner product. We find in this paper,
that for the model investigated, the reality conditions suffice to pick out a
``unique'' inner product on the space of physical states.
Another point of this paper is to avoid the use of any technique specific to
two dimensions, like that of bosonization.

\section{Hamiltonian formulation}
Our starting point is the pseudo-classical Lagrangian density\footnote{We use
signature $(+,-)$ i.e. $\eta _{00}=-\eta _{11}=1$, $[\gamma ^{\mu} ,
\gamma ^{\nu} ]_+=2\eta ^{\mu \nu}$,  with $x^0\equiv t$  (time) and $x^1\equiv
x\in \: [\! -\pi r,\pi r[$  (spatial coordinate
on a circle with radius r). }
\begin{equation}
{\cal L}=-\frac{1}{4}F_{\mu \nu}F^{\mu \nu}+\hbar \overline{\psi} \gamma ^\mu
(i\partial _\mu -e A_\mu) \psi -m\overline{\psi} \psi \label{eq:lagr}
\end{equation}
(throughout we set $c=1$ but keep $\hbar$ for the sake of amusement), where
\begin{eqnarray}
F_{\mu \nu}=\partial _\mu A_\nu-\partial _\nu A_\mu \\
\overline{\psi} = \psi^* \gamma ^0
\end{eqnarray}
and e and m are the coupling and mass constants respectively while $*$ denotes
complex conjugation. $A_\mu$ is the EM-field and $\psi$ is the
(Grassmann odd) fermion field which both
are taken to be $2\pi r$-periodic. Furthermore we define $\gamma = \gamma ^0
\gamma ^1$. An
explicit representation of the gamma matrices is given by $\gamma ^0=\sigma
_1$, $\gamma ^1=-i \sigma _2$ and $\gamma =\sigma _3$.
The canonical procedure applied to the Lagrangian defined in \refpa{eq:lagr} is
described in detail in \cite{su:const}. We will perform it here since it is of
interest to us. In what follows, $\alpha ,\beta ,\ldots$, are taken as indices
labeling the different components of the spinors and the $\gamma$-matrices.
Introducing the momenta
\begin{eqnarray}
\pi ^\mu &=& \frac{\partial {\cal L}}{\partial (\partial _0A_\mu )}=-F^{0
\mu}\\
\pi _\alpha &=& \frac{\partial {\cal L}}{\partial (\partial _0\psi _\alpha)}=-i
\hbar \psi ^* _\alpha\\
\pi ^* _\alpha &=& \frac{\partial {\cal L}}{\partial (\partial _0\psi _\alpha
^*)}=0
\end{eqnarray}
The non-vanishing Bose-Fermi brackets are,
\begin{eqnarray}
\{ A_\mu (x),\pi ^\nu (y)\} =& \delta _\mu ^\nu \delta (x-y)\\
\{ \psi _\alpha (x),\pi _\beta (y)\} =& -\delta _{\alpha \beta} \delta (x-y)\\
\{ \psi _\alpha ^* (x),\pi _\beta ^* (y)\} =& -\delta _{\alpha \beta} \delta
(x-y)
\end{eqnarray}
The Bose-Fermi brackets have the following properties:
\begin{eqnarray}
\{ A,B \} &=& -(-1)^{n_A n_B} \{ B,A\} \\
\{ A,B+C \} &=& \{ A,B \} +\{ A,C\} \\
\{ A,BC \} &=& (-1)^{n_A n_B} B\{ A,C\} +\{ A,B\} C\\
\{ AB,C\} &=& (-1)^{n_B n_C} \{ A,C\} B+A\{ B,C\}
\end{eqnarray}
where $n$ is even or odd depending upon whether the function in question is
Grassmann even or odd respectively. We get
the primary constraints:
\begin{eqnarray}
\phi &=& \pi ^0 \approx 0\\
\phi _\alpha &=& \pi _\alpha +i \hbar \psi _\alpha ^* \approx 0\\
\phi _\alpha ^* &=& \pi _\alpha ^* \approx 0
\end{eqnarray}
The Hamiltonian density becomes:
\bea
{\cal H}_c &=&\de _0 A_\mu \pi ^\mu +\de _0 \psi _\alpha \pi _\alpha+\de _0
\psi _\alpha ^* \pi _\alpha ^*-{\cal L}\nonumber \\
&=& \frac{1}{2} E^2-\hbar \psi ^* \gamma (i \de _1 -e A)\psi +m\psi ^* \gamma
^0 \psi \nonumber \\
&&-A_0 \chi +\de _0 A_0 \phi +\de _0 \psi _\alpha \phi _\alpha +\de _0 \psi
_\alpha ^* \phi _\alpha ^*
\eea
where we've introduced $E=\pi ^1$, $A=A_1$ and $\chi =\de _1 E-e \hbar \psi
_\alpha ^* \psi _\alpha$. Hence we get the total
 Hamiltonian $H_t$:
\be
H_t =\int dx {\cal H}+\int dx(u \phi+v_\alpha \phi _\alpha +w_\alpha \phi
_\alpha ^*)
\ee
where
\be
{\cal H}=\frac{1}{2} E^2-\hbar \psi ^* \gamma (i \de _1 -e A)\psi +m\psi ^*
\gamma ^0 \psi-A_0 \chi ,
\ee
and we've introduced Lagrange multipliers $u,v _\alpha$ and $w_\alpha$, the
latter two being Grassmann odd. Moving on to the
constraint analysis we have:
\bea
\dot{\phi} &=&\{ \phi,H_t \} |_{\phi =\phi _\alpha =\phi _\alpha ^* =0} =\chi
\approx 0\\
\dot{\phi }_\alpha &=& \hbar (i \de _1+e A)(\psi ^* \gamma)_\alpha +m(\psi ^*
\gamma ^0)_\alpha +e \hbar
A_0 \psi _\alpha ^* +i \hbar w_\alpha =0 \label{eq:con1}\\
\dot{\phi }_\alpha ^* &=& \hbar (\gamma (i \de _1 -e A)\psi )_\alpha -m (\gamma
^0 \psi )_\alpha -e \hbar A_0 \psi _\alpha
+i \hbar v_\alpha =0 \label{eq:con2}
\eea
which determines $v_\alpha$ and $w_\alpha$. Hence we have single secondary
constraint $\chi \approx 0$. Using \refpa{eq:con1} and \refpa{eq:con2} one
finds $\dot{\chi } =0$. Now let us consider the
constraint algebra. We obviously have $0=\{ \phi ,\phi _\alpha \}=\{ \phi ,
\phi _\alpha ^* \}=\{ \phi ,\chi \}$, and
\bea
\{ \phi _\alpha (x),\phi _\beta ^* (y) \} &=&-i \hbar \delta _{\alpha \beta}
\delta (x-y)\\
\{ \phi _\alpha (x),\chi (y) \} &=& -e\hbar \psi _\alpha ^* (x) \delta (x-y)\\
\{ \phi _\alpha ^* (x),\chi (y)\} &=& e \hbar \psi _\alpha (x) \delta (x-y)
\eea
Letting $\varphi =\chi-i e(\phi _\alpha \psi _\alpha+\psi _\alpha ^* \phi
_\alpha ^*)$ we find that its bracket
with all the other constraints is weakly zero. Hence $\phi$ and $\varphi$ are
first class, while $\phi _\alpha$ and
$\phi _\alpha ^*$ are second class constraints. Thus
\bea
\Delta _{\alpha \beta}(x,y) &=& \left[ \begin{array}{cc}
\{ \phi _\alpha (x),\phi _\beta (y) \} & \{ \phi _\alpha (x),\phi _\beta ^* (y)
\} \\
\{ \phi _\alpha ^* (x),\phi _\beta (y) \} & \{ \phi _\alpha ^* (x),\phi _\beta
^* (y) \}
\end{array} \right] \nonumber \\
&=& -i\hbar \delta (x-y) \left[ \begin{array}{cc}
0 & \delta _{\alpha \beta}\\
\delta _{\alpha \beta} & 0
\end{array} \right]
\eea
Then
\be
\Delta _{\alpha \beta}^{-1}=\frac{i}{\hbar} \delta (x-y) \left[
\begin{array}{cc}
0 & \delta _{\alpha \beta}\\
\delta _{\alpha \beta} & 0
\end{array} \right]
\ee
since
\be
\int dz \Delta _{\alpha \gamma} (x,z) \Delta _{\gamma \beta}
^{-1}(z,y)=\delta(x-y) \left[ \begin{array}{cc}
\delta _{\alpha \beta} & 0\\
0 & \delta _{\alpha \beta}
\end{array} \right]
\ee
Therefore the Dirac bracket between two arbitrary phase space functions $A$ and
$B$ becomes
\bea
\{ A(x),B(y)\} _D &=& \{ A(x),B(y) \} \nn \\
&-& \int dz dw \{ A(x),\left[ \begin{array}{c} \phi _\alpha (z) \\ \phi _\alpha
^* (z) \end{array} \right] ^T \} \Delta
_{\alpha \beta} ^{-1} (z,w) \{ \left[ \begin{array}{c} \phi _\beta (w) \\ \phi
_\beta ^* (w) \end{array} \right]
,B(y) \} \nn \\
&=& \{ A(x),B(y) \} \nn \\
&-& \int dz dw \{ A(x),\phi _\alpha (z)\} \frac{i}{\hbar} \delta (w-z)
\delta _{\alpha \beta} \{ \phi _\beta ^* (w),B(y)\} \nn \\
&-& \int dz dw \{ A(x),\phi _\alpha ^* (z)\} \frac{i}{\hbar} \delta (w-z)
\delta _{\alpha \beta} \{ \phi _\beta
(w),B(y)\} \nn \\
&=& \{ A(x),B(y) \}-\frac{i}{\hbar}\int dz \{ A(x),\phi _\alpha (z) \} \{ \phi
_\alpha ^* (z),B(y)\} \nn \\
&-& \frac{i}{\hbar}\int dz \{ A(x),\phi _\alpha ^* (z) \} \{ \phi _\alpha
(z),B(y)\}
\eea
i.e.
\be
\{ \psi _\alpha (x),\psi _\beta ^* (y) \} _D =\frac{1}{i\hbar} \delta _{\alpha
\beta} \delta (x-y) \label{eq:bracket}
\ee
The bracket in \refpa{eq:bracket} is the only fermionic bracket we'll need
after having imposed the constraints
$\phi _\alpha =\phi _\alpha ^* =0$. Imposing these two constraints we get
\bea
\pi _\alpha &=&-i\hbar \psi _\alpha ^* \\
\pi _\alpha ^* &=&0
\eea
Our first class constraint $\varphi \approx 0$ becomes
\be
\varphi =\de _1 E-e \hbar \psi _\alpha ^* \psi _\alpha =\chi \approx 0
\label{eq:fcconst}
\ee
which is just Gauss' law. Dropping  the constraint $\phi \approx 0$ and instead
considering $\lambda =A_0$ as a Lagrange
multiplier leads to the reduced Hamiltonian
$H=\int dx {\cal H}_{ph}-H_g$ where
\bea
{\cal H}_{ph} &=& \frac {1}{2} E^2-\hbar \psi ^* \gamma (i\de _1 -e A)\psi +m
\psi ^* \gamma ^0 \psi \\
H_g &=& \int dx \lambda  (x) \chi (x)
\eea
The only non-vanishing brackets are (dropping the subscript $D$):
\bea
\{ A(x),E(y) \} &=& \delta (x-y) \\
\{ \psi _\alpha (x), \psi _\beta ^* (y) \} &=& \frac{1}{i\hbar} \delta _{\alpha
\beta} \delta (x-y)
\eea
Let us finally consider infinitesimal gauge transformations, which are
generated by the first class constraint $H_g$.
\bea
\{ A(x),H_g\} &=& -\de _x \lambda  (x)\\
\{ \psi _\alpha (x),H_g\} &=& ie \lambda  (x)\psi _\alpha (x)\\
\{ \psi _\alpha ^* (x),H_g \} &=& -ie \lambda  (x) \psi _\alpha ^* (x)
\eea
Hence finite gauge transformations are of the form:
\be
\left\{ \begin{array}{ccl}
A'(x) &=& A(x)-\de _x \lambda (x)\\
\psi '(x) &=& e^{ie\lambda (x)}\psi (x)\\
{\psi ^* }'(x) &=& e^{-ie \lambda (x)}\psi ^* (x)
\end{array} \right. \label{eq:gauge}
\ee
\subsection{Loop and line observables}
Of particular interest when doing Dirac quantization are the physical, i.e. the
gauge-invariant, functions on phase space
called observables. Let
\be
T(\gamma )=e^{ie \oint _\gamma dx A(x)}
\ee
where $\gamma$ is a closed loop. The only loops on a circle are the ones that
are labeled by the winding
number $n$, i.e. they wind $n$ times around the circle. Hence write
\be
T(n)=e^{ie \oint _n dx A(x)}
\ee
Furthermore let
\be
L_{\alpha \beta}(x,y)=\psi _\alpha ^* (x) e^{ie\int _x ^y dx' A(x')}\psi _\beta
(y)
\ee
Using \refpa{eq:gauge} it is easily shown that both $T$ and $L_{\alpha \beta}$
are gauge-invariant\footnote{
Here we allow both $x$ and $y$ to take any value on the real line.}. The
bracket algebra between observables
takes the form
\bea
\{ T(n),E(x)\} &=&ie \oint _n dx' \delta (x-x') \: T(n)=ien\: T(n) \\
\{ L_{\alpha \beta}(x,y),E(z)\} &=& \psi _\alpha ^* (x)\psi _\beta (y)ie
\int_x^y dx' \delta (x'-z)
\: e^{ie\int _x ^y dx' A(x')} \nn \\ &=& ie \theta (x,y,z) L_{\alpha
\beta}(x,y)
\eea
where
\bea
\theta (x,y,z) &=& \int _x ^y dx' \delta (x'-z)=\frac{1}{2\pi r}\int _x ^y dx'
\sum _k e^{ik(x'-z)} \nn \\
&=& \frac{y-x}{2\pi r}+\frac{1}{2\pi r}\sum _{k \neq 0}
\frac{1}{ik}(e^{ik(y-z)}-e^{ik(x-z)})\label{eq:gfunc}
\eea
and $k$ can take the values $k=0,\pm \frac{1}{r}, \pm \frac{2}{r},\ldots$ .
Also,
\bea
\{ T(n),L_{\alpha \beta} (x,y) \} &=& 0 \\
\{ L_{\alpha \beta}(x,y),L_{\alpha '\beta '} (x',y')\} &=& \frac{1}{i\hbar }
\delta _{\beta \alpha '}\delta (y-x')
L_{\alpha \beta '}(x,y') \nn \\
&-& \frac{1}{i\hbar }\delta _{\alpha \beta '}\delta (x-y') L_{\alpha
'\beta}(x',y)
\eea
It turns out that it is useful to define $U(x,y)=e^{ie \int _x ^y dx' A(x')}$
and hence
\bea
L_{\alpha \beta}(x,y) &=& U(x,y) \psi _\alpha ^* (x)\psi _\beta (y) \nn \\
&=& U(x,y)\frac{1}{2}(\psi _\alpha ^* (x)\psi _\beta (y)-\psi _\beta (y) \psi
_\alpha ^* (x)) \\
\{ U(x,y),E(z) \} &=& ie \theta (x,y,z) U(x,y) \\
\{ U(x,y),U(x',y') \} &=& 0
\eea
We see that we can express the physical part of the Hamiltonian density, ${\cal
H}_{ph}$, in terms of observables:
\be
{\cal H}_{ph}(x) =\frac{1}{2}E^2(x)-i\hbar \lim _{y \rightarrow x} \gamma
_{\alpha \beta} \de _y
L_{\alpha \beta}(x,y)+m \gamma ^0 _{\alpha \beta } L_{\alpha \beta }(x,x)
\label{eq:clham}
\ee
and the first class constraint \refpa{eq:fcconst} as
\be
\chi (x)=\de _x E(x)-\rho (x) \approx 0 \label{eq:gauss}
\ee
where $\rho$ is the charge density $\rho (x)=e\hbar L_{\alpha \alpha }(x,x)$.
Also of interest is the momentum density,
${\cal P}(x)$,
\[ {\cal P}(x)=-i\psi _\alpha ^* (x)(i\de _x -eA(x))\psi _\alpha (x) \]
i.e.
\be
{\cal P}(x)=-i\hbar \lim _{y \rightarrow x} \de _y L_{\alpha
\alpha}(x,y).\label{eq:clmom}
\ee
Note that the electromagnetic field carries no momentum of its own. This is a
consequence of the fact that pure electromagnetism
in $1+1$-dimensions has no local degrees of freedom.
\section{Quantization}
The general quantization program is described in \cite{rt:diss} and
\cite{aa:lect}.
We quantize the bracket algebra found in the previous section according to the
rule
\be
[\hat{A},\hat{B} ]=i \hbar \widehat{ \{ A,B \} }
\ee
for bosonic functions and
\be
[\hat{A},\hat{B} ]_{_+} =i \hbar \widehat{ \{ A,B \} }
\ee
for fermionic.
Hence quantization consists in finding a representation of the following
commutator algebra:\footnote{We denote commutators by
$\lbrack \: ,\:  \rbrack$ and anticommutators by $\lbrack \: ,\: \rbrack
_{_+}$.}
\bea
[\hat T (n),\hat E (x)] &=& -e\hbar n \hat T (n) \label{eq:loopre}\\
\lbrack \hat L_{\alpha \beta }(x,y),\hat E(z)\rbrack &=& -e\hbar \theta (x,y,z)
\hat L_{\alpha \beta } (x,y) \label{eq:loopl}\\
\lbrack \hat L_{\alpha \beta }(x,y),\hat L_{\alpha '\beta '}(x',y')\rbrack &=&
\delta _{\beta \alpha '}\delta (y-x')
\hat L_{\alpha \beta '}(x,y') \nn \\
&-& \delta _{\alpha \beta '}\delta (x-y') \hat L_{\alpha '\beta}(x',y)
\label{eq:llalg}
\eea
It turns out however, that it is difficult to find a ``good''\footnote{In the
sense of having a representation space consisting
only of gauge-invariant states.} representation of the above observable
algebra, so it seems that we have to represent non gauge-invariant operators as
well (see however the end of this section).
Instead of representing only the
gauge-invariant
operator $\hat L_{\alpha \beta }$ we represent $\hat \psi ,\hat \psi ^*$ and
$\hat U$ also. (The
representation, of a similar algebra, found in \cite{rs:loopre} doesn't seem to
be completely satisfactory). Hence
\bea
\lbrack \hat \psi _\alpha (x),\hat \psi _\beta ^* (y) \rbrack _{_+} &=& \delta
_{\alpha \beta} \delta (x-y) \label{eq:ferma} \\
\lbrack \hat U(x,y),\hat E(z) \rbrack &=& -e\hbar \theta (x,y,z) \hat U(x,y)
\label{eq:partr}
\eea
should also be satisfied by the representation\footnote{We've only written the
non-vanishing commutators and
anticommutators, the vanishing ones should obviously also be satisfied.}. Then
we define
\be
\hat L_{\alpha \beta} (x,y)=\frac{1}{2}\lbrack \hat \psi _\alpha ^* (x),\hat
\psi _\beta (y)\rbrack \hat U(x,y)
\label{eq:lab}
\ee
Let us start out by representing the fermionic operators in \refpa{eq:ferma}.
We use the representation first used in
\cite{fj:func}. Hence let these operators act on wave functionals $\Psi$ of a
complex Grassmann field $\eta (x)$ by:
\bea
\pa{\hat \psi _\alpha (x)  \Psi} (\eta ,\eta ^* ) &=& \frac{1}{\sqrt{2}} (\eta
_\alpha (x)+\frac{\delta}{\delta \eta _\alpha ^* (x)})
\Psi (\eta ,\eta ^* )\\
\pa{\hat \psi _\alpha ^* (x)  \Psi}(\eta ,\eta ^* ) &=& \frac{1}{\sqrt{2}}
(\eta _\alpha ^*(x)+\frac{\delta}{\delta \eta _\alpha (x)})
\Psi (\eta ,\eta ^* )
\eea
Proceeding to \refpa{eq:loopre} we use the loop representation, introduced for
gravity in \cite{rs:loopsp}
and for the pure Maxwell field in \cite{ar:loopma}. We also have to consider
\refpa{eq:partr} simultaneously.
Let the operators $\hat T,\hat E$ and $\hat U$ act on wave functionals of loops
(an integer $n$ in this case), and any
 number of coordinate pairs $(x,y)$ representing the start and endpoints of a
line. Then,
\bea
\pa{\hat T(n)  \Psi} (m,x_1,y_1,\ldots ,x_a,y_a) &=& \Psi (n+m,x_1,y_1,\ldots
,x_a,y_a) \\
\pa{\hat E(x)  \Psi} (m,x_1,y_1,\ldots ,x_a,y_a) &=& -e\hbar (n+\sum _{i=1} ^a
\theta (x_i,y_i,x) ) \times \nn \\
&& \Psi (m,x_1,y_1,\ldots ,x_a,y_a) \\
\pa{\hat U(x,y)  \Psi} (m,x_1,y_1,\ldots ,x_a,y_a) &=& \Psi (m,x_1,y_1,\ldots
,x_a,y_a,x,y)
\eea
We also have
\be
\lbrack \hat U(x,y),\hat U(x',y') \rbrack =0 \label{eq:uualg}
\ee
and classically the $U$:s satisfy $U(x,z)U(z,y)=U(x,y)$, $U(x,x)=1$ and \[
U(x,y+2\pi rm)=U(x-2\pi rm,y)=U(x,y)T(m).\] We want these
identities to be satisfied also quantum-mechanically. Hence by imposing the
following restrictions on the wave functionals,
\bea
\Psi (n,x_1,y_1,\ldots x_i,y_i, \ldots , x_j,y_j,\ldots ,x_a,y_a) &=&
\Psi (n,x_1,y_1,\ldots x_j,y_j, \ldots , x_i,y_i,\ldots ,x_a,y_a) \nn  \\
\Psi (n,x_1,y_1,\ldots ,x_a,y_a,x,z,z,y) &=& \Psi (n,x_1,y_1,\ldots
,x_a,y_a,x,y) \nn \\
\Psi (n,x_1,y_1,\ldots ,x_a,y_a,x,x) &=& \Psi (n,x_1,y_1,\ldots ,x_a,y_a) \nn
\\
\Psi (n,,x_1,y_1,\ldots ,x_a,y_a+2\pi rm) &=&\Psi (n+m,x_1,y_1,\ldots
,x_a,y_a), \label{eq:re1}
\eea
\refpa{eq:uualg} is satisfied and $\hat U(x,z)\hat U(z,y)=\hat U(x,y)$, $\hat
U(x,x)=\hat 1$, \[ \hat U(x,y+2\pi rm)=\hat U(x-2\pi rm,y)=
\hat U(x,y) \hat T(m).\] Combining the two
representations defined, i.e. writing $\Psi (n,x_1,y_1,\ldots ,x_a,y_a,\eta
,\eta ^* )$ and defining
$\hat L_{\alpha \beta}$ by \refpa{eq:lab} one can easily check that
\refpa{eq:loopre}-\refpa{eq:llalg}
are satisfied,  e.g. \refpa{eq:loopre} (suppressing irrelevant
coordinates)\footnote{There are also terms $\eh \sum \theta (x_i,y_i,x)$ which
trivially cancel, hence we don't bother to write them.}:
\bea
\pa{\lbrack \hat T(n),\hat E(x) \rbrack  \Psi} (m) &=& \pa{\hat T(n) \hat E(x)
\Psi} (m)-\pa{\hat E(x)\hat T(n)
 \Psi} (m) \nn \\
&=& \pa{\hat T(n)  \Psi '} (m)-\pa{\hat E(x) \Psi ''} (m)\nn \\
&=& \Psi '(m+n)-(-e\hbar m)\Psi ''(m) \nn \\
&=& -e\hbar (m+n)\Psi (m+n)+e\hbar m \Psi (m+n) \nn \\
&=& -e\hbar n\Psi(n+m) =-e\hbar n\pa{\hat T(n) \Psi} (m)
\eea
where $\Psi '(m)=-e\hbar m\Psi (m)$ and $\Psi ''(m)=\Psi(m+n)$.
The explicit action of $\hat L_{\alpha \beta}$ on a state is (by
\refpa{eq:lab}):
\bea
\lefteqn{\pa{\hat L_{\alpha \beta}(x,y) \Psi}(n,x_1,y_1,\ldots x_a,y_a,\eta
,\eta ^* )=} \nn \\
&& \frac{1}{2}(\eta _\alpha ^* (x)\eta _\beta (y)+
\eta _\alpha ^* (x)\frac{\delta}{\delta \eta _\beta ^* (y)}-\eta _\beta
(y)\frac{\delta}{\delta \eta _\alpha (x)}+
\frac{\delta}{\delta \eta _\alpha (x)}\frac{\delta}{\delta \eta _\beta ^*
(y)})\times \nn \\
&& \Psi (n,x_1,y_1,\ldots x_a,y_a,x,y, \eta,\eta ^* )
\eea

\subsection{Pure Maxwell}
An instructive example on the use of loop representations is the quantization
of pure Maxwell i.e. we have the
Hamiltonian operator
\be
\hat H=\frac{1}{2}\int dx \hat E^2(x)
\ee
and wave functionals of loops $\Psi (n)$. We see that the Gauss' law constraint
is trivially satisfied on loop states
\be
\pa{\de _x \hat E(x)  \Psi} (n)=\de _x (-e\hbar n) \Psi (n) =0
\ee
Letting the Hamiltonian act one finds
\be
\pa{\hat H \Psi} (n)=\frac{1}{2}\int dx (-e\hbar n)^2 \Psi (n)=\pi r(e\hbar )^2
n^2 \Psi (n)
\ee
The eigenstates of $\hat H$ are obviously the characteristic states $\Psi
_m(n)=\delta _{m,n}$ with eigenvalues
$E_m=\pi r (e\hbar m)^2$, $m=0,\pm 1,\pm 2,\ldots$ . To find an inner product
on this vector space we first
consider the classical reality conditions:
\bea
E^* (x) &=& E(x) \\
T^* (n) &=& ( e^{ie \oint _n dx A(x)} )^* =T(-n)
\eea
which, when quantized, turn into the hermitian adjoint relations:
\bea
\hat E^\dagger (x) &=& \hat E(x) \label{eq:ad1}\\
\hat T^\dagger (n) &=& \hat T(-n)\label{eq:ad2}
\eea
We use these relations, \refpa{eq:ad1} and \refpa{eq:ad2}, to find an inner
product. Try the following
ansatz
\be
<\Phi,\Psi > =\sum _{n,n'=-\infty}^\infty \mu (n,n')\Phi^* (n) \Psi (n')
\ee
where the measure $\mu$ is positive definite and satisfies $\mu (n,n') =\mu
(n',n)^* $. Hence $\mu$ should
be chosen such that
\bea
<\Phi,\hat E(x) \Psi > &=& <\hat E(x)\Phi,\Psi > \label{eq:ein}\\
<\Phi,\hat T(m) \Psi > &=& <\hat T(-m) \Phi,\Psi > \label{eq:tin}
\eea
are satisfied. \refpa{eq:ein} leads to
\be
\mu (n,n')=\mu (n)\delta _{n,n'} \label{eq:mu1}
\ee
and \refpa{eq:tin} implies
\be
\mu (n,n'-m)=\mu (n+m,n') \label{eq:mu2}
\ee
Combining \refpa{eq:mu1} and \refpa{eq:mu2} we get
\be
\mu (n)=\mu (n+m), \: \: \forall m
\ee
and thus $\mu (n,n')=C \delta _{n,n'}$ for some real constant $C$, choose $C=1$
for definiteness, i.e.
\be
<\Phi,\Psi >=\sum _{n=-\infty}^{\infty}\Phi^* (n)\Psi  (n) \label{eq:innerpr}
\ee
which also is positive definite. We see that the reality conditions suffice to
pick out a ``unique'' inner product on
the state space. Note further that the characteristic states $\Psi _m$ form an
ON-basis for the Hilbert space defined
using the inner product \refpa{eq:innerpr}.

\subsection{Gauge-invariant states}
Returning to the interacting theory we've come to the subject of
gauge-invariant states. The gauge-invariant states
are at the heart of Dirac quantization, all other states are considered
irrelevant. The classical Gauss' law constraint
\refpa{eq:gauss}, when quantized should annihilate physical states i.e.
\be
(\de _x \hat E(x)-\hat \rho (x)) \Psi _{phys}=0 \label{eq:phys}
\ee
where $\hat \rho (x)=\eh \epl \hat L_{\alpha \alpha} (x,x+\ep )$. The idea to
construct physical states is the following: first find one state,
then by acting upon this state by the various physical operators we can create
new physical states. This first state should be as simple as possible.
Call this state $\psb$. Assume that it factorizes as:
\be
\psb (n,x_1,y_1,\ldots ,x_a,y_a,\eta ,\eta ^* )=\psb (n,x_1,y_1,\ldots
,x_a,y_a) \psb (\eta ,\eta ^* )
\ee
We solve \refpa{eq:phys} for $\psb$ by having it satisfy:
\bea
\de _x \he (x) \psb &=&0 \label{eq:ephys}\\
\hr (x) \psb &=&0 \label{eq:rphys}
\eea
\refpa{eq:ephys} implies
\bea
\lefteqn{ \pa{\de _x \he (x) \psb }(n,x_1,y_1,\ldots ,x_a,y_a,\eta ,\eta ^* )
=} \nn \\
&& -\eh \de _x(n+\sum _{i=1}^a \theta (x_i,y_i,x)) \psb (n,x_1,y_1,\ldots
,x_a,y_a,\eta ,\eta ^* )= \nn \\
&& -\eh \sum _{i=1}^a (\delta (x-x_i)-\delta (x-y_i)) \psb (n,x_1,y_1,\ldots
,x_a,y_a,\eta ,\eta ^* ) =0
\eea
i.e.
\be
\sum _{i=1}^a (\delta (x-x_i)-\delta (x-y_i)) \psb (n,x_1,y_1,\ldots
,x_a,y_a)=0.
\ee
Hence $\psb (n,x_1,y_1)$ can only have support on $x_1+2\pi rm=y_1$ for any
integer $m$. Therefore let $\psb $ satisfy:
\be
\psb (n,x_1,y_1)=\left\{ \begin{array}{ll}
\psb (n+m) & \mbox{if $x_1+2\pi rm=y_1$} \\
0 & \mbox{otherwise}
\end{array} \right.
\ee
and $\psb (n,x_1,y_1,\ldots ,x_a,y_a)$ is non-zero only if it can be reduced to
$\psb (n+m,x_1,x_1)$ by using
\refpa{eq:re1}. Any such state can be written as a superposition of states
$\psb ^{n_0}$ where $\psb ^{n_0}(n)=\delta _{n_0,n}$.
Proceeding to \refpa{eq:rphys} we make an ansatz for $\psb (\eta ,\eta ^* )$:
\be
\psb (\eta ,\eta ^* )=\exp (\int dy \: \eta _\beta ^* (y) M_{\beta \gamma}\eta
_\gamma (y))
\ee
where $M_{\beta \gamma}$ is a constant matrix. Hence we get (being very careful
with the anticommutativity
of $\eta$ ):
\bea
\lefteqn{ \epl \pa{\hat L_{\alpha \alpha}(x,x+\ep ) \psb } (n,x_1,y_1,\ldots
,x_a,y_a,\eta ,\eta ^* )=} \nn \\
&& \frac{1}{2} \epl (\eta _\alpha ^* (x) \eta _\alpha (x+\ep )+M_{\alpha
\alpha} \delta (\ep )-
\eta _\alpha ^* (x)M_{\alpha \beta}M_{\beta \gamma}\eta _\gamma (x+\ep ))
\times \nn \\
&& \psb (n,x_1,y_1,\ldots ,x_a,y_a,x,x+\ep ,\eta ,\eta ^* )=0. \label{eq:gainv}
\eea
To have a well-defined limit we demand $M_{\alpha \alpha}=0$ i.e. tr$M=0$.
Having done this we see that if
$M^2=1$, then \refpa{eq:gainv} is satisfied. Thus $\psb$ is gauge-invariant
provided $M$ is a linear
combination $a\gamma ^0+ib\gamma ^1+c\gamma$ such that $a^2+b^2+c^2=1$. Let us
now consider some properties of
$\psb ^{n_0}$. A calculation gives:
\be
\hat E(z) \prod _{i=1}^a \hat U(x_i,y_i) \psb ^{n_0}=-\eh (n_0+\sum_{i=1}^a
\theta (y_i,x_i,z)) \prod _{j=1}^a \hat U(x_j,y_j)
\psb ^{n_0}.\label{eq:uuu}
\ee
We will show this relation in the simplest case. First note (suppressing
coordinates irrelevant for the calculation and for simplicity
of notation restricting all coordinates to $[\! -\pi r,\pi r[$):
\bea
\lefteqn{\pa{\hat U(x_1,y_1) \psb ^{n_0}}(n,x,y)=\psb^{n_0} (n,x,y,x_1,y_1)=}
\nn \\
&& \left\{ \begin{array}{ll}
\delta _{n_0,n} & \mbox{if ($x=y$ and $x_1=y_1$) or ($x=y_1$ and $y=x_1$)}\\
0 & \mbox{otherwise}
\end{array} \right.
\eea
Hence
\bea
\pa{\hat E(z)\hat U(x_1,y_1) \psb^{n_0}}(n,x,y) &=& -\eh (n+\theta (x,y,z))
\psb^{n_0} (n,x,y,x_1,y_1)\nn \\
&=& -\eh (n_0+\theta (y_1,x_1,z)) \psb ^{n_0}(n,x,y,x_1,y_1)\nn \\
&=& -\eh (n_0+\theta (y_1,x_1,z))\pa{\hat U(x_1,y_1) \psb ^{n_0}}(n,x,y)
\eea
since $\theta (x,x,z)=0$. To completely specify the state $\psb^{n_0}$ let us
make the choices $M=\pm i\gamma ^1$ (for
reasons that will become apparent in a while). We can always, by suitable
redefinitions of the gamma
matrices, transform into those cases (we can obviously transform from $M=i
\gamma ^1$ to $M=-i\gamma ^1$
as well but we choose not to). Now, acting with the physical operator
$A_{\alpha \beta} \hat L_{\alpha \beta}$ on $\psb^{n_0}$
we get a new gauge-invariant state (A is a matrix):
\bea
\pa{A_{\alpha \beta} \hat L_{\alpha \beta}(x,y) \psb ^{n_0} }(n,x_1,y_1,\ldots
,x_a,y_a,\eta ,\eta ^* )=\nn \\
(\eta _\alpha ^* (x)\tilde A_{\alpha \beta} \eta _\beta (y)
+\frac{1}{2}\mbox{tr}(MA) \delta(x-y))
 \psb ^{n_0} (n,x_1,y_1,\ldots ,x_a,y_a,x,y,\eta ,\eta ^* ) \label{eq:lineb}
\eea
where $\tilde A=\frac{1}{2} (1-M)A(1+M)$. The tilde map has a number of
important properties:
\be
\begin{array}{ll}
(1-M)\tilde A =2\tilde A & \tilde A (1+M)=2\tilde A \\
\tilde A (1-M)=0 & (1+M)\tilde A =0 \\
\tilde A_1 \tilde A_2 =0 & \mbox{tr}\tilde A =0\\
\tilde 1=0 & \tilde M=0,
\end{array} \label{eq:tilde1}
\ee
in general, and
\be
\begin{array}{ll}
\tilde \gamma ^1=0 & \tilde \gamma ^0 =\pm i \tilde \gamma \\
\frac{1}{2}(\gt \gamma \gt )=\gt & \frac{1}{2}\mbox{tr}(\gamma \gt )=1\\
\frac{1}{2}(\gt \gamma ^0 \gt )=\mp i \gt & \frac{1}{2}\mbox{tr}(\gamma ^0 \gt
)=\mp i
\end{array} \label{eq:tilde2}
\ee
when $M=\pm i\gamma ^1$. We see that the vector space of tilded matrices is
one-dimensional and we choose $\gt$ as a basis for these matrices. Now define:
\be
\Psi ^+_{(n_0)} =\psb ^{n_0}
\ee
using $M=i \gamma ^1$ i.e.
\bea
\lefteqn{\Psi ^+_{(n_0)}(n,x_1,y_1,\ldots ,x_a,y_a,\eta ,\eta ^* )=} \nn \\
&&  \psb ^{n_0} (n,x_1,y_1,\ldots ,x_a,y_a)
\exp (\int dx \eta _\alpha ^* (x) i \gamma ^1 _{\alpha \beta} \eta _\beta (x))
\eea
and similarly for $\Psi ^-_{(n_0)}$ using $M=-i\gamma ^1$. Inspired by the form
of \refpa{eq:lineb}
we further define:
\bea
\lefteqn{\Psi ^+_{(n_0,x_1,y_1,\ldots ,x_a,y_a)} (n,x'_1,y'_1,\ldots
x'_b,y'_b,\eta ,\eta ^* )=}\nn \\
&&  \psb ^{n_0} (n,x_1,y_1,\ldots x_a,y_a,x'_1,y'_1,\ldots ,x'_b,y'_b) \times
\nn \\
&& \eta _{\alpha _1}^* (x_1)\gt _{\alpha _1 \beta _1} \eta _{\beta _1} (y_1)
\cdots \eta _{\alpha _a}^* (x_a)
\gt _{\alpha _a \beta _a} \eta _{\beta _a} (y_a) \times \nn \\
&& \exp (\int dx \eta _\alpha ^* (x) i \gamma ^1 _{\alpha \beta} \eta _\beta
(x))
\eea
using $M=+i\gamma ^1$ in the $\gt$:s, and similarly define $\Psi
^-_{(n_0,x_1,y_1,\ldots ,x_a,y_a)}$ using $M=-i\gamma ^1$
everywhere. By \refpa{eq:uuu} these states have the property:
\bea
\hat E(z) \Psi ^+_{(n,x_1,y_1,\ldots ,x_a,y_a)}=-\eh (n+\sum _{i=1}^a\theta
(y_i,x_i,z)) \Psi ^+_{(n,x_1,y_1,\ldots ,x_a,y_a)} \label{eq:eplus}\\
\hat E(z) \Psi ^-_{(n,x_1,y_1,\ldots ,x_a,y_a)}=-\eh (n+\sum _{i=1}^a\theta
(y_i,x_i,z)) \Psi ^-_{(n,x_1,y_1,\ldots ,x_a,y_a)}. \label{eq:eminus}
\eea
Checking gauge-invariance according to \refpa{eq:phys}, they're indeed found to
be gauge-invariant. Thus we've
found a lot of physical states. Are they complete in the sense that they
suffice to define an inner product on the
space of physical states? To find an inner product we have to know the reality
conditions of all operators acting on this space.
The relevant operators are $\hat E$, $\hat T(m)$ and $\hat L _{\alpha \beta}$.
As in the previous section we have
$\hat E^\dagger (x)=\hat E(x)$ and $\hat T^\dagger (m)=\hat T(-m)$.
Classically, $L_{\alpha \beta}$ satisfy:
\be
L_{\alpha \beta}^* (x,y)=L_{\beta \alpha}(y,x).
\ee
Since a basis for all $2\times 2$ matrices is given by $1,\gamma ^0,\gamma ^1$
and $\gamma$ we get a basis for all
$\hat L_{\alpha \beta}$:s by contracting with these matrices. Noting that
$1,\gamma ^0$ and $\gamma$ are hermitian matrices while
$\gamma ^1$ is anti-hermitian we get the hermitian adjoint relations:
\be
\begin{array}{ll}
\lone ^\dagger (x,y)=\lone (y,x) & \lgn ^\dagger (x,y)=\lgn (y,x) \\
\lgo ^\dagger (x,y)=-\lgo (y,x) & \lgg ^\dagger (x,y)=\lgg (y,x).
\end{array} \label{eq:herm}
\ee
It is not clear, at this stage, whether all these relations can be implemented
simultaneously. That remains to be seen.
Write ${\cal P}=\sum _M \oplus {\cal P}^M$, where ${\cal P}^M$ denotes the
space of states of the type
$\ppl$ having a fixed $M$. As a shorthand, write ${\cal P}^+={\cal P}^{+i\gamma
^1}$ and
${\cal P}^-={\cal P}^{-i\gamma ^1}$. As we will show by construction, physical
operators map ${\cal P}^M$ into itself i.e. it forms an
invariant subspace in the space of physical states ${\cal P}$. We will
call the subspaces ${\cal P}^M$ the different sectors of the theory. A general
state in ${\cal P}^+$ can be written
\[ \sum _{n=-\infty}^{\infty} \Psi _{f_n}^+, \] where, (all integrals are taken
from $-\pi r$ to $\pi r$),
\be
\Psi _{f_n}^+=f^{(0)}_n\Psi ^+ +\sum _{a=1}^\infty \frac{1}{a!}\int d^a\! x \,
d^a\! y \, f^{(a)}_n(x_1,y_1,\ldots ,x_a,y_a)
\Psi ^+_{(n,x_1,y_1,\ldots ,x_a,y_a)},\label{eq:genst}
\ee
and the functions $f^{(a)}_n$ are required to have the symmetry property:
\be
f^{(a)}_n(x_1,y_1,\ldots ,x_i,y_i,\ldots ,x_j,y_j,\ldots
,x_a,y_a)=f^{(a)}_n(x_1,y_1,\ldots ,x_j,y_j,\ldots ,x_i,y_i,\ldots ,x_a,y_a).
\ee
Similarly define $\Psi _{f_n}^-$ in ${\cal P}^-$. By straightforward but
tedious calculations, using \refpa{eq:tilde1} and \refpa{eq:tilde2},
 we've found the action of the various $\hat L$ operators on these states. For
${\cal P}^+$ we have, (restricting $x$ and $y$ to $[\! -\pi r,\pi r[$),
\bea
\lefteqn{\lone (x,y) \Psi _{f_n}^+ =} \nn \\
&& \sum _{a=1}^\infty \frac{1}{a!}\int d^a\! x \, d^a\! y \, \sum _{b=1}^a \{
\delta (x-x_b) f_n^{(a)}(x_1,y_1,\ldots ,y,y_b,\ldots ,x_a,y_a)-\nn \\
&& \delta (y-y_b)f_n^{(a)}(x_1,y_1,\ldots ,x_b,x,\ldots ,x_a,y_a)\} \ppl
\label{eq:oneact}\\
\lefteqn{\lgn (x,y) \Psi _{f_n}^+ =-i f_n^{(1)}(y,x) \Psi ^+_{(n)} +} \nn \\
&& i\sum _{a=1}^\infty \frac{1}{a!}\int d^a\! x \, d^a\! y \, \{ \sum _{b=1}^a
\{  f_n^{(a-1)}(x_1,y_1,\ldots ,x_b\leave ,y_b \leave ,\ldots ,x_a,y_a) \times
\nn \\
&& \delta (x-x_b)\delta (y-y_b) \}- f_n^{(a+1)}(x_1,y_1,\ldots
,x_a,y_a,y,x)+\nn \\
&& \sum _{b=1}^a f_n^{(a+1)}(x_1,y_1,\ldots ,x_b,x,y,y_b,\ldots ,x_a,y_a)\}
\ppl \label{eq:gnact}\\
\lefteqn{\lgo (x,y) \Psi _{f_n}^+ =-i\delta (x-y) \Psi _{f_n}^+ +}\nn \\
&& i\sum _{a=1}^\infty \frac{1}{a!}\int d^a\! x \, d^a\! y \, \sum _{b=1}^a \{
\delta (x-x_b)
f_n^{(a)}(x_1,y_1,\ldots ,y,y_b,\ldots ,x_a,y_a)+\nn \\
&& \delta (y-y_b)f_n^{(a)}(x_1,y_1,\ldots ,x_b,x,\ldots ,x_a,y_a) \} \ppl \\
\lefteqn{\lgg (x,y) \Psi _{f_n}^+ =f_n^{(1)}(y,x) \Psi ^+_{(n)} +} \nn \\
&& \sum _{a=1}^\infty \frac{1}{a!}\int d^a\! x \, d^a\! y \, \{ \sum _{b=1}^a
\{  f_n^{(a-1)}(x_1,y_1,\ldots ,x_b\leave ,y_b \leave ,\ldots ,x_a,y_a) \times
\nn \\
&& \delta (x-x_b)\delta (y-y_b) \}+ f_n^{(a+1)}(x_1,y_1,\ldots
,x_a,y_a,y,x)-\nn \\
&& \sum _{b=1}^a f_n^{(a+1)}(x_1,y_1,\ldots ,x_b,x,y,y_b,\ldots ,x_a,y_a)\}
\ppl \label{eq:ggact}
\eea
where $/$ denotes the absence of the indicated argument. Furthermore, by
\refpa{eq:eplus} we have
\bea
\lefteqn{\hat E(x) \Psi _{f_n}^+ =}\nn \\
&& -\eh \sum _{a=1}^\infty \frac{1}{a!}\int d^a\! x \, d^a\! y \, \sum _{b=1}^a
\theta (y_b,x_b,x) f_n^{(a)}(x_1,y_1,\ldots ,x_a,y_a)
\times \nn \\ && \ppl .
\label{eq:epd}
\eea
Also,
\bea
\lefteqn{\hat T(m)\Psi _{f_n}^+=}\nn \\
&& f_n^{(0)}\Psi ^+_{(n-m)} +\sum _{a=1}^\infty \frac{1}{a!}\int d^a\! x \,
d^a\! y \, f_n^{(a)}(x_1,y_1,\ldots ,x_a,y_a)
\Psi ^+_{(n-m,x_1,y_1,\ldots ,x_a,y_a)}.
\eea
The corresponding expressions for ${\cal P}^-$ are found just by exchanging $i
\leftrightarrow -i$ in the appropriate formula, hence:
\bea
\lefteqn{\lone (x,y) \Psi _{f_n}^- =} \nn \\
&& \sum _{a=1}^\infty \frac{1}{a!}\int d^a\! x \, d^a\! y \, \sum _{b=1}^a \{
\delta (x-x_b) f_n^{(a)}(x_1,y_1,\ldots ,y,y_b,\ldots ,x_a,y_a)-\nn \\
&& \delta (y-y_b)f_n^{(a)}(x_1,y_1,\ldots ,x_b,x,\ldots ,x_a,y_a)\} \pmi \\
\lefteqn{\lgn (x,y) \Psi _{f_n}^- =i f_n^{(1)}(y,x) \Psi ^-_{(n)} -} \nn \\
&& i\sum _{a=1}^\infty \frac{1}{a!}\int d^a\! x \, d^a\! y \, \{ \sum _{b=1}^a
\{  f_n^{(a-1)}(x_1,y_1,\ldots ,x_b\leave ,y_b \leave ,\ldots ,x_a,y_a) \times
\nn \\
&& \delta (x-x_b)\delta (y-y_b) \}- f_n^{(a+1)}(x_1,y_1,\ldots
,x_a,y_a,y,x)+\nn \\
&& \sum _{b=1}^a f_n^{(a+1)}(x_1,y_1,\ldots ,x_b,x,y,y_b,\ldots ,x_a,y_a)\}
\pmi \\
\lefteqn{\lgo (x,y) \Psi _{f_n}^- =i\delta (x-y) \Psi _{f_n}^- -}\nn \\
&& i\sum _{a=1}^\infty \frac{1}{a!}\int d^a\! x \, d^a\! y \, \sum _{b=1}^a \{
\delta (x-x_b)
f_n^{(a)}(x_1,y_1,\ldots ,y,y_b,\ldots ,x_a,y_a)+\nn \\
&& \delta (y-y_b)f_n^{(a)}(x_1,y_1,\ldots ,x_b,x,\ldots ,x_a,y_a) \} \pmi \\
\lefteqn{\lgg (x,y) \Psi _{f_n}^- =f_n^{(1)}(y,x) \Psi ^-_{(n)} +} \nn \\
&& \sum _{a=1}^\infty \frac{1}{a!}\int d^a\! x \, d^a\! y \, \{ \sum _{b=1}^a
\{  f_n^{(a-1)}(x_1,y_1,\ldots ,x_b\leave ,y_b \leave ,\ldots ,x_a,y_a) \times
\nn \\
&& \delta (x-x_b)\delta (y-y_b) \}+ f_n^{(a+1)}(x_1,y_1,\ldots
,x_a,y_a,y,x)-\nn \\
&& \sum _{b=1}^a f_n^{(a+1)}(x_1,y_1,\ldots ,x_b,x,y,y_b,\ldots ,x_a,y_a)\}
\pmi
\eea
and
\bea
\lefteqn{\hat E(x) \Psi _{f_n}^- =}\nn \\
&& -\eh \sum _{a=1}^\infty \frac{1}{a!}\int d^a\! x \, d^a\! y \, \sum _{b=1}^a
\theta (y_b,x_b,x) f_n^{(a)}(x_1,y_1,\ldots ,x_a,y_a)
\times \nn \\ && \pmi ,
\eea
\bea
\lefteqn{\hat T(m)\Psi _{f_n}^-=}\nn \\
&& f_n^{(0)}\Psi ^-_{(n-m)} +\sum _{a=1}^\infty \frac{1}{a!}\int d^a\! x \,
d^a\! y \, f_n^{(a)}(x_1,y_1,\ldots ,x_a,y_a)
\Psi ^-_{(n-m,x_1,y_1,\ldots ,x_a,y_a)}.
\eea
Thus we see explicitly that the various sectors ${\cal P}^+$, ${\cal
P}^-,\ldots$ are invariant under the action
of the gauge-invariant operators. Now we can proceed to define an inner
product. It is natural to require that the different
sectors are orthogonal e.g.
\be
<\Psi ^+,\Psi ^->=0\label{eq:orth}
\ee
Then for the $+$ sector we make the ansatz, using the experience from the last
section, ($\Psi _{g_m}^+$ is defined similarly to
$\Psi _{f_n}^+$):
\bea
\lefteqn{<\Psi _{g_m}^+,\Psi _{f_n}^+>=}\nn \\
&& \delta _{n,m}\sum _{a,b} \int d^a\! x \, d^a\! y \, d^b\! x' \, d^b\! y' \,
\mu (x_1,y_1,\ldots ,x_a,y_a,x'_1,y'_1,\ldots ,x'_b,y'_b)
\times \nn \\
&& {g ^{(a)}_m}^* (x_1,y_1,\ldots ,x_a,y_a) f_n^{(b)}(x'_1,y'_1,\ldots
,x'_b,y'_b).
\eea
Imposing $\hat E^\da (x)=\hat E(x)$ using \refpa{eq:epd} we get
\bea
\lefteqn{\mu (x_1,y_1,\ldots ,x_a,y_a,x'_1,y'_1,\ldots ,x'_b,y'_b)=}\nn \\
&& \delta _{a,b} \, \mu (x_1,y_1,\ldots ,x_a,y_a,x'_1,y'_1,\ldots ,x'_a,y'_a).
\eea
Imposing $\lgg ^\da (x,y)=\lgg (y,x)$, i.e.
\[ <\Psi _{g_m}^+,\lgg (x,y) \Psi _{f_n}^+>=<\lgg (y,x) \Psi _{g_m}^+,\Psi
_{f_n}^+>,\]
we find by a tedious calculation using \refpa{eq:ggact} that the measure $\mu$
is fully determined (apart from an
arbitrary overall normalization factor), e.g.
\[ \mu (x_1,y_1,x'_1,y'_1)=\delta (x_1-x'_1) \delta (y_1-y'_1) \]
and hence:
\bea
\lefteqn{<\Psi _{g_m}^+,\Psi _{f_n}^+>=\delta _{n,m}( {g_n^{(0)}}^*
f_n^{(0)}+}\nn \\
&& \sum _{a=1}^\infty \frac{1}{a!} \int d^a\! x \, d^a\! y \, \ep _{i_1\cdots
i_a} {g _n^{(a)}}^* (x_1,y_{i_1},\ldots ,
x_a,y_{i_a})\times \nn \\
&& f_n^{(a)}(x_1,y_1,\ldots ,x_a,y_a)) \label{eq:inner}
\eea
where $\ep _{i_1\cdots i_a}$ is the totally antisymmetric symbol. Checking the
remaining adjoint relations in
\refpa{eq:herm} we find that they are all satisfied by the inner product
\refpa{eq:inner}. Similarly for the $-$ sector
we get an analogous result:
\bea
\lefteqn{<\Psi _{g_m}^-,\Psi _{f_n}^->=\delta _{n,m} ( {g_n^{(0)}}^*
f_n^{(0)}+}\nn \\
&& \sum _{a=1}^\infty \frac{1}{a!} \int d^a\! x \, d^a\! y \, \ep _{i_1\cdots
i_a} {g_n ^{(a)}}^* (x_1,y_{i_1},\ldots ,
x_a,y_{i_a})\times \nn \\
&& f_n^{(a)}(x_1,y_1,\ldots ,x_a,y_a)).
\eea

\subsection{Cyclic representation}
Having, with much labour, calculated all gauge-invariant states in the
functional representation it is easy to see that that
these states can be constructed in a much simpler, algebraic manner. Introduce
the operators,
\[
\begin{array}{ll}
\hat L_+(x,y)=\frac{1}{2}(\gamma-i \gamma ^0)_{\alpha \beta} \hat L_{\alpha
\beta}(x,y) &
\hat L_-(x,y)=\frac{1}{2}(\gamma+i \gamma ^0)_{\alpha \beta} \hat L_{\alpha
\beta}(x,y)\\
\hat K_+(x,y)=\frac{1}{2}(1-i \gamma ^1)_{\alpha \beta} \hat L_{\alpha
\beta}(x,y) &
\hat K_-(x,y)=\frac{1}{2}(1+i \gamma ^1)_{\alpha \beta} \hat L_{\alpha
\beta}(x,y).
\end{array} \]
Using \refpa{eq:llalg}, we obtain,
\bea
\lbrack \lp (x,y),\lp (x',y') \rbrack &=& \lbrack \lm (x,y),\lm (x',y') \rbrack
=0\nn \\
\lbrack \lm (x,y),\lp (x',y') \rbrack &=& \delta (y-x') \km (x,y')-\delta
(x-y')\kp (x',y)\nn \\
\lbrack \km (x,y),\km (x',y') \rbrack &=& \delta (y-x')\km (x,y')-\delta
(x-y')\km (x',y)\nn \\
\lbrack \kp (x,y),\kp (x',y') \rbrack &=& \delta (y-x')\kp (x,y')-\delta
(x-y')\kp (x',y)\nn \\
\lbrack \kp (x,y),\km (x',y') \rbrack &=& 0\nn \\
\lbrack \km (x,y),\lm (x',y') \rbrack &=& \delta (x'-y)\lm (x,y')\nn \\
\lbrack \km (x,y),\lp (x',y') \rbrack &=& -\delta (x-y')\lp (x',y)\nn \\
\lbrack \kp (x,y),\lp (x',y') \rbrack &=& \delta (x'-y)\lp (x,y')\nn \\
\lbrack \kp (x,y),\lm (x',y') \rbrack &=& -\delta (x-y')\lm
(x',y).\label{eq:alg1}
\eea
Also, by \refpa{eq:loopre} and \refpa{eq:loopl},
\bea
\lbrack \hat T(n),\hat E(x) \rbrack &=&-\eh n \hat T(n)\nn \\
\lbrack \hat L(x,y),\hat E(z) \rbrack &=&-\eh \theta (x,y,z) \hat
L(x,y),\label{eq:alg2}
\eea
where in the last equation, $\hat L$ can stand for any one of the operators
$\lp$, $\lm$, $\kp$ or $\km$. Furthermore, we
have the relations,
\[ \hat T(n) \hat T(m)=\hat T(n+m) \]
and
\[ \hat L(x,y+2\pi rm)=\hat L(x-2\pi rm,y)=\hat L(x,y) \hat T(m).\]
Now introduce the cyclic states $\Psi ^+_{(0)}$ and $\Psi ^-_{(0)}$ with the
properties
\[
\begin{array}{ll}
<\Psi ^+_{(0)},\Psi ^+_{(0)} >=<\Psi ^-_{(0)},\Psi ^-_{(0)}>=1, & <\Psi
^+_{(0)},\Psi ^-_{(0)} >=0.
\end{array} \]
In what follows, restrict $x$ and $y$ to the interval $[\! -\pi r,\pi r[$,
\[ \begin{array}{llll}
\hat E(x) \Psi ^+_{(0)} =0, & \lm (x,y)\Psi ^+_{(0)} =0, & \hat T(m) \Psi
^+_{(0)}=\Psi ^+_{(-m)}, & \lp (x,y) \Psi ^+_{(0)} =
\Psi ^+_{(0,x,y)}\\
\hat E(x) \Psi ^-_{(0)} =0, & \lp (x,y)\Psi ^-_{(0)} =0, & \hat T(m) \Psi
^-_{(0)}=\Psi ^-_{(-m)}, & \lm (x,y) \Psi ^-_{(0)} =
\Psi ^-_{(0,x,y)} \end{array} \]
\[ \begin{array}{ll}
\kp (x,y) \Psi ^+_{(0)}=-\frac{1}{2}\delta (x-y)\Psi ^+_{(0)}, & \km (x,y) \Psi
^+_{(0)}=\frac{1}{2}\delta (x-y)\Psi ^+_{(0)} \\
\km (x,y) \Psi ^-_{(0)}=-\frac{1}{2}\delta (x-y)\Psi ^-_{(0)}, & \kp (x,y) \Psi
^-_{(0)}=\frac{1}{2}\delta (x-y)\Psi ^-_{(0)}.
\end{array} \]
Furthermore,
\[ \hat T(-n) \prod _{j=1}^a\lp (x_j,y_j) \Psi ^+_{(0)}=\ppl , \]
\[ \hat T(-n) \prod _{j=1}^a\lm (x_j,y_j) \Psi ^-_{(0)}=\pmi ,\]
and
\[ \begin{array}{lll}
\lp ^\dagger (x,y)=\lm (y,x), & \kp ^\dagger (x,y) =\kp (y,x), & \km ^\dagger
(x,y) =\km (y,x) ,\\
\hat T^\dagger (n)=\hat T(-n), & \hat E^\dagger (x)=\hat E(x).
\end{array} \]
Using the algebra, defined in \refpa{eq:alg1} and \refpa{eq:alg2}, we can
easily reproduce all results found in the
previous subsection, including the inner product.

\section{Dynamics}
Until now, it hasn't really been important what the Hamiltonian looks like. The
discussion in the previous section about
physical states applies to any theory with a Hamiltonian having the same
symmetries (and the same field content), i.e. being
constrained by Gauss' law \refpa{eq:phys}. But we're interested in the specific
Hamiltonian which is given classically by
\refpa{eq:clham} and hence we write
\be
\hat H=\intl{x}(\frac{1}{2} \hat E^2(x)-i\hbar \epl \frac{\de}{\de (x+\ep
)}\lgg (x,x+\ep )+m \epl \lgn (x,x+\ep )). \label{eq:qham}
\ee
$\hat H$ contains the two operators $\lgg (x,x+\ep )$ and $\lgn (x,x+\ep )$. If
we let these operators act on the state $\psb ^{n_0}$
defined in the previous section we see, by \refpa{eq:lineb}, that if we are to
have a finite limit when $\ep$ tends to zero, we
must demand tr$(M\gamma )=0$ and tr$(M\gamma ^0)=0$. These two conditions,
together with the previously found conditions tr$M=0$
and $M^2=1$, implies that $M=\pm i\gamma ^1$ (which is the reason why we've
already studied these cases). Hence, of all
the sectors of physical states available, we rule out all except for the $+$
and $-$ sector when considering the specific dynamics
governed by the Hamiltonian \refpa{eq:qham}, i.e. physical states are in
general a superposition of states from the $+$ and $-$ sector.
To solve for the dynamics, we essentially have to find a complete set of
eigenstates to the Hamiltonian.
The momentum operator, which is given classically by \refpa{eq:clmom}, is
\be
\hat P=-i\hbar \intl{x}\epl \lone (x,x+\ep ).\label{qmom}
\ee
Since $\hat H$ and $\hat P$ commute, they can be simultaneously diagonalized.
Thus we label states by their energy and momentum
eigenvalues. Let us try to find the ground state. Denote it by $\Omega$ and
write $\Omega=C_+ \Omega^+ +C_-
\Omega^-$ where $\Omega^+$
and $\Omega^-$ belongs to the $+$ and $-$ sector respectively ($C_+$ and $C_-$
are constants). The ground state satisfies:
\bea
\hat H\Omega &=& E_{(0)} \Omega \label{eq:hvac}\\
\hat P\Omega &=& 0,\label{eq:pvac}
\eea
where $E_{(0)}$ denotes the vacuum energy. Note that there is no way we can
``normal order'' the Hamiltonian prior to solving for $\Omega$
to get rid of $E_{(0)}$. We simply have to live with the vacuum energy. After
having solved for $\Omega$ we can effectively discard it. The appearance
of $E_{(0)}$ in \refpa{eq:hvac} implies that not only the ground state
satisfies \refpa{eq:hvac} and \refpa{eq:pvac}, but also a whole
range of other states. We will see further on how to pick out the correct
one(s). We make the ansatz
\bea
\Omega _{n_0}^+=\Psi _{G_+}^+\\
\Omega _{n_0}^-=\Psi _{G_-}^-
\eea
with support on $n=n_0$ and where $G_+^{(0)}=G_-^{(0)}=1$, i.e. (by
\refpa{eq:genst})
\bea
\Omega _{n_0}^+ = \Psi _{(n_0)}^+ +\sum _{a=1}^\infty \frac{1}{a!}\int d^a\! x
\, d^a\! y \, G_+^{(a)}(x_1,y_1,\ldots ,x_a,y_a)
\pplo \\
\Omega _{n_0}^- = \Psi _{(n_0)}^- +\sum _{a=1}^\infty \frac{1}{a!}\int d^a\! x
\, d^a\! y \, G_-^{(a)}(x_1,y_1,\ldots ,x_a,y_a))
\pmio .
\eea
Now we're ready to consider \refpa{eq:hvac} for $\Omega _{n_0}^+$ and $\Omega
_{n_0}^-$. We will first do a simple case in detail.
\[ \hat E(x)\Psi ^+_{(n_0,x_1,y_1)}=-\eh (n_0+\theta (y_1,x_1,x))\Psi
^+_{(n_0,x_1,y_1)} \]
and hence,
\[ \hat E(x)^2 \Psi ^+_{(n_0,x_1,y_1)}=(\eh )^2 (n_0^2+2n_0 \theta
(y_1,x_1,x)+\theta (y_1,x_1,x)^2)\Psi ^+_{(n_0,x_1,y_1)} ,\]
i.e.
\[ \frac{1}{2}\intl{x}\hat E^2(x)\Psi ^+_{(n_0,x_1,y_1)} =\frac{(\eh )^2}{2} (2
\pi r n_0^2 -2n_0 (y_1-x_1)+
V(x_1,y_1))\Psi ^+_{(n_0,x_1,y_1)}\]
where the potential $V$ is given by (using \refpa{eq:gfunc})
\bea
V(x_1,y_1) &=& \intl{x}\theta (x_1,y_1,x)^2=\frac{(y_1-x_1)^2}{2\pi
r}+p(x_1-y_1) \\
p(x_1-y_1) &=& \frac{1}{\pi r} \sum _{k\neq 0}\frac{1}{k^2}(1-e^{ik(x_1-y_1)}).
\eea
Furthermore define
\be
W_{n_0}(x_1,y_1)=V(x_1,y_1)-2n_0(y_1-x_1).
\ee
Note that $V(x_1,y_1)=|x_1-y_1|$ as long as $-\pi r<x_1,y_1<\pi r$. In general
we get
\bea
\lefteqn{\frac{1}{2}\intl{x}\hat E^2(x)\Omega _{n_0}^+=\frac{(\eh )^2}{2}\{
2\pi rn_0^2\Psi _{(n_0)}^++}\nn \\
&& \sum _{a=1}^\infty \frac{1}{a!}\int d^a\! x \, d^a\! y \, (2\pi r
n_0^2+W_{n_0}(x_1,y_1,\ldots ,x_a,y_a))\times \nn \\
&& G_+^{(a)}(x_1,y_1,\ldots ,x_a,y_a) \pplo \}
\eea
where
\be
W_{n_0}(x_1,y_1,\ldots ,x_a,y_a)=V(x_1,y_1,\ldots ,x_a,y_a)-2n_0(y_1-x_1+\cdots
+y_a-x_a)\label{eq:wn0}
\ee
and
\bea
V(x_1,y_1,\ldots ,x_a,y_a) &=& \frac{1}{2\pi r}(y_1-x_1+\cdots +y_a-x_a)^2+\sum
_{i,j=1}^ap(x_i-y_j)\nn \\
&-& \sum _{j>i=1}^a(p(x_i-x_j)+p(y_i-y_j)).
\eea
Hence, using \refpa{eq:gnact} and \refpa{eq:ggact}, we get
\bea
\lefteqn{\hat H \Omega _{n_0}^+=( (\eh )^2\pi rn_0^2-i\intl{x}\epl (\hbar \de
_{x+\ep }+m)G_+^{(1)}(x+\ep ,x))
\Psi _{(n_0)}^++}\nn \\
&& \sum _{a=1}^\infty \frac{1}{a!}\int d^a\! x \, d^a\! y \, \{ \frac{(\eh
)^2}{2}(2\pi rn_0^2 +
W_{n_0}(x_1,y_1,\ldots ,x_a,y_a))\times \nn \\
&& G_+^{(a)}(x_1,y_1,\ldots ,x_a,y_a)+\nn \\
&& i \sum _{b=1}^a G_+^{(a-1)}(x_1,y_1,\ldots ,x_b\leave ,y_b\leave ,\ldots
,x_a,y_a)(\hbar \frac{\de}{\de y_b}+m)
\delta (y_b-x_b)-\nn \\
&& i \intl{x}\epl (\hbar \de _{x+\ep }+m)G_+^{(a+1)}(x_1,y_1,\ldots
,x_a,y_a,x+\ep ,x)+\nn \\
&& i \intl{x}\epl (\hbar \de _{x+\ep }+m)\sum
_{b=1}^aG_+^{(a+1)}(x_1,y_1,\ldots ,x_b,x,x+\ep ,y_b,\ldots ,x_a,y_a)\}
\times \nn \\
&& \pplo =E_{(0)} \Omega _{n_0}^+. \label{eq:Theeqn}
\eea
Thus we can identify,
\be
E_{(0)}=\pi r(\eh n_0)^2-i\intl{x}\epl (\hbar \de _{x+\ep }+m)G_+^{(1)}(x+\ep
,x)\label{eq:ve}
\ee
and
\bea
\lefteqn{-i\intl{x}\epl (\hbar \de _{x+\ep }+m)\sum
_{b=1}^aG_+^{(a+1)}(x_1,y_1,\ldots ,x_b,x,x+\ep ,
y_b,\ldots ,x_a,y_a)+}\nn \\
&& i\intl{x}\epl (\hbar \de _{x+\ep }+m)G_+^{(a+1)}(x_1,y_1,\ldots
,x_a,y_a,x+\ep ,x)=\nn \\
&& i \intl{x}\epl (\hbar \de _{x+\ep }+m)G_+^{(1)}(x+\ep
,x)G_+^{(a)}(x_1,y_1,\ldots ,x_a,y_a)-\nn \\
&& i\sum_{b=1}^aG_+^{(a-1)}(x_1,y_1,\ldots ,x_b\leave ,y_b\leave ,\ldots
,x_a,y_a)(\hbar \de _{x_b}-m)
\delta (x_b-y_b)+\nn \\
&& \frac{(\eh )^2}{2}W_{n_0}(x_1,y_1,\ldots ,x_a,y_a) G_+^{(a)}(x_1,y_1,\ldots
,x_a,y_a).\label{eq:v1}
\eea
Continuing on to \refpa{eq:pvac} and using \refpa{eq:oneact} we find
\bea
\lefteqn{\hat P\Omega _{n_0}^+=}\nn \\
&& -i\hbar \Psi _{n_0}\sum _{a=1}^\infty \frac{1}{a!}\int d^a\! x \, d^a\! y \,
\sum_{b=1}^a
(\de _{x_b}+\de _{y_b})G_+^{(a)}(x_1,y_1,\ldots ,x_a,y_a)\times \nn \\
&& \pplo =0\nn
\eea
i.e.
\[ \sum_{b=1}^a(\de _{x_b}+\de _{y_b})G_+^{(a)}(x_1,y_1,\ldots ,x_a,y_a)=0 \]
and hence $G_+^{(a)}$ is translationally invariant, i.e.
\be
G_+^{(a)}(x_1+d,y_1+d,\ldots ,x_a+d,y_a+d)=G_+^{(a)}(x_1,y_1,\ldots
,x_a,y_a)\label{eq:v3}
\ee
for any displacement $d$. All in all, \refpa{eq:ve}-\refpa{eq:v3} determine
the $+$ sector component of the ground state.\refpa{eq:ve} and \refpa{eq:v1}
are not really well defined as
they stand, because they involve the evaluation of a distribution at zero.
Hence, we take it as being
understood, that an ultraviolet cutoff has been performed to regularize these
equations. Similarly,
the corresponding expressions for the $-$ sector are found by letting
$m\leftrightarrow -m$. Thus,
\be
E_{(0)}=\pi r(\eh n_0)^2-i\intl{x}\epl (\hbar \de _{x+\ep }-m)G_-^{(1)}(x+\ep
,x),\label{eq:ve-}
\ee
\bea
\lefteqn{-i\intl{x}\epl (\hbar \de _{x+\ep }-m)\sum
_{b=1}^aG_-^{(a+1)}(x_1,y_1,\ldots ,x_b,x,x+\ep ,
y_b,\ldots ,x_a,y_a)+}\nn \\
&& i\intl{x}\epl (\hbar \de _{x+\ep }-m)G_-^{(a+1)}(x_1,y_1,\ldots
,x_a,y_a,x+\ep ,x)=\nn \\
&& i\intl{x}\epl (\hbar \de _{x+\ep }-m)G_-^{(1)}(x+\ep
,x)G_-^{(a)}(x_1,y_1,\ldots ,x_a,y_a)-\nn \\
&& i\sum_{b=1}^aG_-^{(a-1)}(x_1,y_1,\ldots ,x_b\leave ,y_b\leave ,\ldots
,x_a,y_a)(\hbar \de _{x_b}+m)
\delta (x_b-y_b)+\nn \\
&& \frac{(\eh )^2}{2}W_{n_0}(x_1,y_1,\ldots ,x_a,y_a) G_-^{(a)}(x_1,y_1,\ldots
,x_a,y_a)\label{eq:v2-}
\eea
and
\be
G_-^{(a)}(x_1+d,y_1+d,\ldots ,x_a+d,y_a+d)=G_-^{(a)}(x_1,y_1,\ldots
,x_a,y_a).\label{eq:v3-}
\ee
Hence we see that $\Omega _{n_0}^+$ and $\Omega _{n_0}^-$ are coupled only
through the vacuum energy $E_{(0)}$.

\subsection{Vacuum and VEV's when $e=0$}
Before embarking on a solution attempt to the interacting vacuum equations
\refpa{eq:ve}-\refpa{eq:v3-}, let us study the
special case $e=0$. Then the potential term drops out from \refpa{eq:v1} and
\refpa{eq:v2-}. Write
$g_+(x_1,y_1)=G_+^{(1)}(x_1,y_1)$ and $g_-(x_1,y_1)=G_-^{(1)}(x_1,y_1)$. It is
easily seen that a solution to all the
vacuum equations is given as (independently of $n_0$),
\bea
g_+(x_1,y_1)=-\frac{1}{2\pi r}\sum _ke^{ik(x_1-y_1)}\frac{(\hbar k+im)}{E_k}\\
g_-(x_1,y_1)=-\frac{1}{2\pi r}\sum _ke^{ik(x_1-y_1)}\frac{(\hbar k-im)}{E_k}
\eea
where $E_k=\sqrt{(\hbar k)^2+m^2}$ and
\bea
G_+^{(a)}(x_1,y_1,\ldots ,x_a,y_a) &=& g_+(x_1,y_1)g_+(x_2,y_2)\cdots
g_+(x_a,y_a)\\
G_-^{(a)}(x_1,y_1,\ldots ,x_a,y_a) &=& g_-(x_1,y_1)g_-(x_2,y_2)\cdots
g_-(x_a,y_a),
\eea
i.e.
\bea
\Omega _{n_0}^+ &=& \exp{ ( \intl{x_1} d\! y_1 \, g_+ (x_1,y_1)\lp
(x_1,y_1))}\Psi _{(n_0)}^+\nn \\
\Omega _{n_0}^- &=& \exp{ ( \intl{x_1} d\! y_1 \, g_- (x_1,y_1)\lm
(x_1,y_1))}\Psi _{(n_0)}^-.\label{eq:om}
\eea
The true groundstate when $e=0$ should be independent of the electromagnetic
degrees of freedom, hence consider the states
\[
\Omega ^+=\sum _{n_0=-\infty}^\infty \Omega _{n_0}^+, \]
and
\[
\Omega ^-=\sum _{n_0=-\infty}^\infty \Omega _{n_0}^- .\]

This solution really is the ground state in the case $e=0$ even though we will
postpone the proof of this until later. The
expression for $E_{(0)}$ is, for both the $+$ and $-$ sector using
\refpa{eq:ve} and \refpa{eq:ve-},
\[ E_{(0)}=-\sum _k E_k . \]
The cutoff is performed by allowing Fourier modes up to $|k|=\frac{N}{r}$ where
$N$ is some integer. Note the
following identities
\bea
\intl{x}g_+^* (x,y)g_+(x,y') &=& \frac{1}{2\pi r}\sum
_ke^{ik(y-y')}=\delta(y-y')\\
\intl{x}g_-^* (x,y)g_-(x,y') &=& \delta(y-y')
\eea
and
\be
g_+^* (x_1,y_1)=g_-(y_1,x_1).
\ee
Also,
\[ \intl{x_1} d\! y_1 \, g_+^* (x_1,y_1)g_+(x_1,y_1)=2 \pi r\delta
(0)=2N+1=\lambda , \]
where we've introduced $\lambda =2N+1$.
We further note that the cutoff delta functions work exactly like ordinary
delta functions provided they are
integrated against functions that are similarly cut off. With this knowledge,
we proceed to calculate some
vacuum expectation values. Hence using \refpa{eq:inner}, (dropping the sum over
$n_0$ which just factors out),
\bea
\lefteqn{<\Omega^+,\Omega^+>=1+\sum _{a=1}^\infty \frac{1}{a!}\int d^a\! x \,
d^a\! y \,
\ep _{i_1\cdots i_a} g_+^* (x_1,y_{i_1})\cdots g_+^* (x_a,y_{i_a})\times} \nn
\\
&& g_+(x_1,y_1) \cdots g_+(x_a,y_a)=\nn \\
&& 1+\sum _{a=1}^\infty \frac{1}{a!}\int d^a\! y \, \ep _{i_1\cdots i_a}
\delta (y_{i_1}-y_1) \cdots \delta (y_{i_a}-y_a)=\nn \\
&& 1+\sum _{a=1}^\infty p_a(\lambda )=2^\lambda \, \, \, \, (\lambda \geq 0),
\eea
where
\be
p_a(\lambda )=\frac{1}{a!}\prod _{i=0}^{a-1}(\lambda -i).
\ee
In a similar manner
\[ <\Omega^-,\Omega^->=2^\lambda .\]
Hence, by \refpa{eq:orth},
\be
<\Omega,\Omega>=2^\lambda (|C_+|^2+|C_-|^2)\label{eq:vacex}
\ee
where $\Omega=C_+ \Omega^++C_- \Omega^-$.
Next let us consider the expectation value of the operator $\lgg (x,y)$. Using
\refpa{eq:ggact} we get
\bea
\lefteqn{<\Omega^+,\lgg (x,y)\Omega^+>=g_+(y,x) 2^\lambda +}\nn \\
&& (g_-(y,x)-g_+(y,x)) \sum _{a=1}^{\infty } a \frac{p_a(\lambda )}{\lambda
}=\nn \\
&& g_+(y,x) 2^\lambda +(g_-(y,x)-g_+(y,x))2^{\lambda -1} \, \, \, (\lambda \geq
1).
\eea
Similarly,
\bea
\lefteqn{<\Omega^-,\lgg (x,y)\Omega^->=}\nn \\
&& g_-(y,x) 2^\lambda +(g_+(y,x)-g_-(y,x))2^{\lambda -1} \, \, \, (\lambda \geq
1).
\eea
Hence,
\bea
\lefteqn{<\Omega,\lgg (x,y)\Omega> = |C_+|^2 (g_+(y,x) 2^\lambda
+(g_-(y,x)-g_+(y,x))2^{\lambda -1})+}\nn \\
&& |C_-|^2 (g_-(y,x) 2^\lambda +(g_+(y,x)-g_-(y,x))2^{\lambda -1}).\hspace{3cm}
\eea
Something very odd has happened. By \refpa{eq:herm} we should have
\[ (\lgg (x,y))^\da =\lgg (y,x) \] i.e.
\[ <\Omega,\lgg (x,y)\Omega>^* =<\Omega,\lgg (y,x)\Omega> ,\]
but
\bea
\lefteqn{<\Omega,\lgg (x,y)\Omega>^* = |C_+|^2 (g_-(x,y) 2^\lambda
+(g_+(x,y)-g_-(x,y))2^{\lambda -1})+}\nn \\
&& |C_-|^2 (g_+(x,y) 2^\lambda +(g_-(x,y)-g_+(x,y))2^{\lambda -1}),\hspace{3cm}
\eea
since $g_+^* (x,y)=g_-(y,x)$. Hence for consistency we must demand
$|C_+|=|C_-|$ i.e. $C_+$ and $C_-$ are equal up to an
irrelevant phase. Thus, choosing
\[ C_+=C_-=1 \]
i.e. $\Omega=\Omega^++\Omega^-$, we find
\be
<\Omega,\lgg (x,y)\Omega> = (g_+(y,x)+g_-(y,x)) 2^\lambda ,
\ee
and hence, using \refpa{eq:vacex},
\bea
\frac{<\Omega,\lgg (x,y)\Omega>}{<\Omega,\Omega>} &=&
\frac{1}{2}(g_+(y,x)+g_-(y,x))\nn \\
&=& -\frac{1}{2\pi r}\sum _k e^{ik(y-x)}\frac{\hbar k}{E_k}.
\eea
We note that the end result is independent of the cutoff $N$, as it should be.
What has happened? In a previous subsection
we constructed an inner product that was to have all the correct reality
properties, yet when we use it here it doesn't have
the right properties unless $|C_+|=|C_-|$. The only reasonable explanation of
this mystery, is that somehow the distributional
character of $g_+$ and $g_-$ with its regularization, ruined the (formal)
derivation. One can also argue that $\Omega ^+$ and
$\Omega ^-$ are improper states (not normalizable) and we cannot expect the
adjoint relations to hold on arbitrary such states.
 Furthermore, we note that something had to go
wrong, since otherwise we wouldn't have obtained a definite answer for the
expectation value. At this stage we can only
speculate about what happens when $e\neq 0$, but the apparent symmetry between
the $+$ and $-$ sector seems to suggest that
this is a general result, i.e. $\Omega _{n_0}=\Omega _{n_0}^++\Omega _{n_0}^-$
even when $e\neq 0$. Similarly, we  find
\bea
\frac{<\Omega,\lgn (x,y)\Omega>}{<\Omega,\Omega>} &=&
\frac{i}{2}(g_-(y,x)-g_+(y,x))\nn \\
&=& -\frac{1}{2\pi r}\sum _k e^{ik(y-x)}\frac{m}{E_k}
\eea
and
\bea
\frac{<\Omega,\lgo (x,y)\Omega>}{<\Omega,\Omega>} &=&0\\
\frac{<\Omega,\lone (x,y)\Omega>}{<\Omega,\Omega>} &=&0.
\eea
All these results agree with
\be
\bra{0} \frac{1}{2}[\hat \psi _\alpha ^\da (x),\hat \psi _\beta (y)]
\ket{0}=-\frac{1}{4\pi r}
\sum _k e^{ik(y-x)}\frac{(\hbar k\gamma +m\gamma ^0)_{\alpha \beta}}{E_k}
\ee
obtained using ordinary free quantization of fermions. This is encouraging.
Even though we have required gauge invariance
in all cases, i.e. even in the case $e=0$, we get the same result as the one
one gets having set $e=0$ from the outset
and then of course not worrying about gauge invariance. Note also that had we
calculated the expectation value of
these operators on the state $\Omega _{n_0}$, we would have obtained the same
results as above, as long as $x$ and $y$
 were in the interval $[\! -\pi r,\pi r[$. For values of $x$ and $y$ outside
this interval the expectation value is zero for
all $\hat L$ operators.

\subsection{Solution attempt to the vacuum equations}
Let us try to solve \refpa{eq:ve}-\refpa{eq:v3-} in the general case. The
solution should, in the limit $e\rightarrow 0$
approach the solution found in the previous section. This is sufficient for the
solution to be the vacuum (or a vacuum in the
case of vacuum degeneracy). By looking at the form of the potential $W_{n_0}$
in \refpa{eq:wn0} and ignoring the two
non-periodic terms that depends on $y_1-x_1+\ldots y_a-x_a$, we make the ansatz
(being for the moment only concerned
about the $+$ sector),
\be
G_+^{(a)}(x_1,y_1,\ldots ,x_a,y_a)=g_+(x_1,y_1)\cdots g_+(x_a,y_a) \phi
^{(a)}(x_1,y_1,\ldots ,x_a,y_a)
\ee
where
\be
\phi ^{(a)}(x_1,y_1,\ldots ,x_a,y_a)=\exp \{ \sum _{i,j=1}^a\varphi (x_i-y_j)-
\sum _{j>i=1}^a (\varphi (x_i-x_j)+\varphi (y_i-y_j))\} ,
\ee
and we demand $\varphi (0)=0$, $\varphi (x)=\varphi (-x)$ which implies
$\varphi '(0)=0$. Furthermore, we let
$\varphi$ be a $2\pi r$-periodic function.
We have not been able to generalize this ansatz (in any good way) to
accommodate a functional dependence on
$y_1-x_1+\ldots $. We will anyway insert it into the vacuum equations, just
dropping the non-periodic terms in $W_{n_0}$.
This might be a valid short distance approximation. We note that the ansatz
satisfies \refpa{eq:v3}.
Let us discuss some properties of the function $\phi ^{(a)}$. We have
\bea
\phi ^{(a+1)}(x_1,y_1,\ldots ,x_a,y_a,x,x) &=& \phi ^{(a)}(x_1,y_1,\ldots
,x_a,y_a)\\
\phi ^{(a+1)}(x_1,y_1,\ldots ,x_b,x,x,y_b,\ldots ,x_a,y_a) &=& \phi
^{(a)}(x_1,y_1,\ldots ,x_a,y_a)
\eea
and
\bea
\lefteqn{\epl \de _{x+\ep } \phi ^{(a+1)}(x_1,y_1,\ldots ,x_a,y_a,x+\ep
,x)=}\nn \\
&& \phi ^{(a)}(x_1,y_1,\ldots ,x_a,y_a) \sum _{i=1}^a (\varphi '(x-y_i)-\varphi
'(x-x_i))\\
\lefteqn{\epl \de _{x+\ep } \phi ^{(a+1)}(x_1,y_1,\ldots ,x_b,x,x+\ep
,y_b,\ldots ,x_a,y_a)=}\nn \\
&& \phi ^{(a)}(x_1,y_1,\ldots ,x_a,y_a) \sum _{i=1}^a (\varphi '(x-y_i)-\varphi
'(x-x_i)).
\eea
Let us next consider \refpa{eq:v1}. Using the properties of the $e=0$ solution
we find that \refpa{eq:v1} becomes
\bea
\lefteqn{-i\hbar \sum _{b=1}^a \prod _{j \neq b}g_+(x_j,y_j) \{ \delta
(y_b-x_b) \sum _{i=1}^a (\varphi '(y_i-y_b)-
\varphi '(x_i-y_b))+}\nn \\
&& \intl{x}g_+(x_b,x)g_+(x,y_b)\sum _{i=1}^a(\varphi '(x-y_i)-\varphi
'(x-x_i))\} =\nn \\
&& \frac{(\eh )^2}{2}W_{n_0}(x_1,y_1,\ldots ,x_a,y_a)g_+(x_1,y_1)\cdots
g_+(x_a,y_a).\hspace{3cm}\label{eq:anz}
\eea
When $a=1$, \refpa{eq:anz} is, (dropping the non-periodic terms in $W_{n_0}$),
\be
-i\hbar \intl{x}g_+(x_1,x)g_+(x,y_1)(\varphi '(x-y_1)-\varphi '(x-x_1))=\frac{
(\eh )^2}{2}p(x_1-y_1)g_+(x_1,y_1),
\label{eq:anz1}
\ee
and when $a=2$ we have, using \refpa{eq:anz1},
\bea
\lefteqn{-i\hbar g_+(x_1,y_1)\intl{x}g_+(x_2,x)g_+(x,y_2)(\varphi
'(x-y_1)-\varphi '(x-x_1))-}\nn \\
&& i \hbar g_+(x_1,y_1)\delta (y_2-x_2) (\varphi '(y_1-y_2)-\varphi
'(x_1-y_2))+(x_1\leftrightarrow x_2,y_1\leftrightarrow y_2)=\nn \\
&& \frac{(\eh )^2}{2}(p(x_1-y_2)+p(x_2-y_1)-p(x_2-x_1)-p(y_2-y_1))\times \nn \\
&& g_+(x_1,y_1)g_+(x_2,y_2).\label{eq:anz2}
\eea
When $x_1=x_2$ and $y_1=y_2$, \refpa{eq:anz2} reduces to \refpa{eq:anz1}.
Furthermore, when $a>2$, \refpa{eq:anz} still
contain terms of the same type as \refpa{eq:anz2}. Hence if \refpa{eq:anz2} is
satisfied, then \refpa{eq:anz} is satisfied
for all $a$. In the case $m=0$, an explicit solution to \refpa{eq:anz1} might
be found. By letting $m\rightarrow 0$ in
the expression for $g_+$, we have
\[ g_+(x_1,y_1)=-\frac{1}{2\pi r}(i+\sum _{k \neq 0}
e^{ik(x_1-y_1)}\mbox{sgn}(k)). \]
Writing
\[ \varphi (x_1-y_1)=\sum _{k\neq 0}a_k (e^{ik(x_1-y_1)}-1), \]
where $a_k=a_{-k}$, we find the following solution to \refpa{eq:anz1}
\be
\varphi (x_1-y_1)=\frac{e^2\hbar}{4\pi r}\sum _{k\neq 0}\frac{
\mbox{sgn}(k)}{k^3}(e^{ik(x_1-y_1)}-1).
\ee
This solution seems to satisfy \refpa{eq:anz2} as well, even though we haven't
really proved this. For a general $m$,
\refpa{eq:anz1} takes the form,
\bea
\lefteqn{\sum _{k' (k-k' \neq 0)}\frac{(\hbar k'+i m)(\hbar k'+im)}{E_k
E_{k'}}(k-k')a_{k-k'}=}\nn \\
&& \frac{e^2 \hbar }{4\pi r}( \sum _{k' (k-k' \neq 0)} \frac{\hbar
k'+im}{E_{k'} (k-k')^2}-r^2 \frac{\pi ^2}{3}
\frac{\hbar k+im}{E_k} )\, \, \, , \forall k .\label{eq:anzz}
\eea
We haven't found an analytic solution to \refpa{eq:anzz} but it can be solved
numerically. It remains to extend the ansatz to
include also the non-periodic terms in $W_{n_0}$. We might get a hint on how to
do this modification by calculating the
ground state using ordinary time-independent perturbation theory. Writing the
Hamiltonian as $\hat H=\hat H_0+\hat H_I$ where
\[
\hat H_0=\intl{x} (-i\hbar \epl \frac{\de}{\de (x+\ep )}\lgg (x,x+\ep )+m \epl
\lgn (x,x+\ep )),
\]
and
\[
\hat H_I=\frac{1}{2}\intl{x}\hat E^2(x) ,
\]
and regarding $H_I$ as a perturbation, perturbation theory proceeds in the
usual manner. Note that the groundstate of $H_I$ is
infinitely degenerate, each one of the states $\Omega _{n_0}$ given by
\refpa{eq:om} being as good as any other. However, to
be able to do this, we need to know all excited states as well
when $e=0$. We will find these states presently.

\subsection{Excited states}
We expect a one-pair state to have the form $\hat a \Omega$ (at least when
$e=0$), where
\be
\hat a=\intl{x} \, dy N_{\alpha \beta}(x,y) \hat L_{\alpha \beta}(x,y),
\ee
and
\[ N_{\alpha \beta}(x,y)=A(x,y)\delta _{\alpha \beta}+B(x,y) \gamma _{\alpha
\beta}+C(x,y) \gamma _{\alpha \beta}^0 +
D(x,y) \gamma _{\alpha \beta}^1 .\]
We want $\hat a \Omega$ to be an eigenstate of $\hat H$, with energy $E'$.
Hence
\be
[ \hat H,\hat a ]\Omega =E \hat a \Omega ,\label{eq:exen}
\ee
where $E=E'-E_{(0)}$. We also want $\hat a \Omega$ to be an eigenstate of $\hat
P$, with momentum $P$, i.e.
\be
[ \hat P,\hat a ] \Omega=P \hat a \Omega  .\label{eq:exmom}
\ee
We obtain,
\bea
\lefteqn{ \lbrack \hat H, \hat a \rbrack =\frac{ (\eh )^2}{2}\intl{x}\, dy
V(x,y) N_{\alpha \beta}(x,y) \hat L_{\alpha \beta}(x,y)+}\nn \\
&& \eh \intl{z} \, dx \, dy \theta (x,y,z)N_{\alpha \beta}(x,y) \hat L_{\alpha
\beta}(x,y) \hat E(z)-\nn \\
&& i\hbar \intl{x} \, dy \{ (\de _x A(x,y)+\de _y A(x,y))\gamma _{\alpha
\beta}+(\de _xB(x,y)+\de _y B(x,y))\delta _{\alpha \beta}+\nn \\
&& (\de _yC(x,y)-\de _xC(x,y))\gamma _{\alpha \beta}^1 +(\de _yD(x,y)-\de _x
D(x,y))\gamma _{\alpha \beta}^0\}\hat L_{\alpha \beta}(x,y)+\nn \\
&& 2m\intl{x}\, dy (B(x,y)\gamma _{\alpha \beta}^1+D(x,y)\gamma _{\alpha
\beta})\hat L_{\alpha \beta}(x,y),
\eea
and
\be
\lbrack \hat P,\hat a \rbrack =-i\hbar \intl{x}\, dy \{ (\de _x+\de
_y)N_{\alpha \beta}(x,y)\} \hat L_{\alpha \beta}(x,y) .
\ee
We realize that if there is any solution to \refpa{eq:exen} when $e\neq 0$,
then it is state dependent. However, when $e=0$, we can find
state independent solutions to \refpa{eq:exen} and \refpa{eq:exmom}. Let us
investigate this in detail. We find,
\be
\begin{array}{ll}
-i\hbar (\de _x+\de _y)A=P A & -i\hbar (\de _x+\de _y)B=PB \\
-i\hbar (\de _x+\de _y)C=PC & -i\hbar (\de _x+\de _y)D=PD \\
-i\hbar (\de _x+\de _y)A+2mD=EB & -i\hbar (\de _x+\de _y)B=EA \\
-i\hbar (\de _y-\de _x)C+2mB=ED & -i\hbar (\de _y-\de _x)D=E C. \label{eq:sind}
\end{array}
\ee
If \refpa{eq:sind} is satisfied, then \refpa{eq:exen} and \refpa{eq:exmom} are
satisfied. Solving \refpa{eq:sind} and demanding $A,B,C$ and $D$
to be $2\pi r$-periodic functions of both $x$ and $y$ separately leads to,
\be
\begin{array}{ll}
A(x,y)=P e^{\frac{i}{\hbar}(p_1 x+p_2 y)} & B(x,y)=E e^{\frac{i}{\hbar}(p_1
x+p_2 y)}\\
C(x,y)=\frac{(E^2-P^2)(p_2-p_1)}{2mE}e^{\frac{i}{\hbar}(p_1 x+p_2 y)} &
D(x,y)=\frac{(E^2-P^2)}{2m}e^{\frac{i}{\hbar}(p_1 x+p_2 y)},
\end{array}
\ee
where
\[
\begin{array}{ll}
P=p_1+p_2 &  E=\pm E_1\pm E_2 \\
E_1=\sqrt{p_1^2+m^2} &  E_2=\sqrt{p_2^2+m^2}
\end{array}
\]
and $p_1=\frac{\hbar n_1}{r}$, $p_2=\frac{\hbar n_2}{r}$ for arbitrary integers
$n_1$ and $n_2$. Hence the one-pair states are labelled by $p_1$ and
$p_2$ and we have recovered the two-particle interpretation of these states
when $e=0$. Acting with the so defined $\hat a(p_1,p_2)$ on the $e=0$
ground state we find,
\bea
\lefteqn{\hat a(p_1,p_2)\Omega =i2\pi r\delta _{p_1,-p_2}\,
\frac{E}{2m}(E+2E_1)(\Omega^- -\Omega^+)+}\nn \\
&& \Psi _{n_0}\frac{(E+E_1+E_2)(E-E_1+E_2)(E+E_1+E_2)}{2E\, E_1\, E_2}\times
\nn \\
&& \sum _{a=1}^\infty \frac{1}{a!} \int d^a\! x \, d^a\! y \, \sum _{b=1}^a
e^{\frac{i}{\hbar}(p_1 x_b+p_2y_b)}\times \nn \\
&& \{ (E_1+E_2+\frac{i}{m}(E_1\, p_2 -E_2\, p_1))\prod _{j\neq b}g_+(x_j,y_j)
\pplo+\nn \\
&& (E_1+E_2-\frac{i}{m}(E_1\, p_2 -E_2\, p_1))\prod _{j\neq b}g_-(x_j,y_j)
\pmio \}.
\eea
Thus we see explicitly that $\hat a$ annihilates $\Omega$ when
\[ \begin{array}{lll}
E=-E_1+E_2, & E=E_1-E_2, & E=-E_1-E_2 ,
\end{array} \]
which proves that we have found the correct ground state when $e=0$. Denote the
corresponding operators by $\hat a _{-+}(p_1,p_2)$,
$\hat a _{+-}(p_1,p_2)$ and $\hat a _{--}(p_1,p_2)$. In the remaining case
$E=E_1+E_2$ the corresponding operator is denoted $a_{++}(p_1,p_2)$
(which creates a pair from vacuum). Also,
\[ \begin{array}{ll} \hat a_{++}^\da (p_1,p_2)=-\hat a_{--}(-p_2,-p_1), & \hat
a_{+-}^\da (p_1,p_2)=-\hat a_{-+}(-p_2,-p_1) .
\end{array} \]
Thus, when $e=0$, an arbitrary excited state can be constructed by acting with
products of $\hat a_{++}$:s on the ground state. We won't go
into the complicated discussion of trying to solve \refpa{eq:exen} when $e\neq
0$ as it requires an explicit knowledge of the interacting
ground state, which we don't have. Note however that \refpa{eq:exmom} is solved
independently of $e$.

\section{Discussion}
As we have seen, Dirac quantization works perfectly well for electrodynamics on
a cylindrical space-time. Not finding the explicit ground state is
a disappointment since comparing calculated expectation values found using our
methods with previous results would have been interesting. However,
we believe that with some more effort, the exact ground state in this formalism
will be found. The formalism is easily extended to $3+1$-dimensions
and it should be interesting to investigate what happens in that case. The
appearance of photons will necessarily complicate the discussion.
Another generalization of this model is Yang-Mills coupled to fermions. The
representation defined in this paper should, without too much effort,
be generalizable to the non-abelian case. Finally, gravity coupled to fermions
is a theory that should be investigated in this framework.

\vspace{10mm}
{\Large{\bf Acknowledgements}}

\vspace{3mm}
We wish to thank Bo-Sture Skagerstam for suggesting the model, and him and
Ingemar Bengtsson for discussions and criticism. Furthermore, we
thank Bengt Nilsson for support and discussions.

\end{document}